\documentstyle[12pt]{article}
\title{Complexity and criticality in Laplacian growth models}
\author{{\bf F. Guinea\cite{csic} and O. Pla} \vspace{0.5cm}\\
The Harrison M. Randall Laboratory of Physics. \\
The University of Michigan.\\
Ann Arbor. Michigan 48109-1120 \vspace{1cm}\\
{\bf E. Louis} \vspace{0.5cm}\\
Departamento de F\'{\i}sica Aplicada. Universidad de Alicante. \\
Apartado 99. 03080 Alicante. Spain.}
\date{}
\begin{document}
\maketitle
\begin{abstract}
We analyze the dynamical evolution of systems which obey simple growth
laws, like diffusion limited aggregation or dielectric breakdown.
We show that, if the developing patterns is sufficiently complex,
a scale invariant noise spectrum is generated, in agreement with the
hypothesis that the system is in a self organized critical state. The
intrinsic noise generated in the evolution is shown to be independent of
the (extrinsic) stochastic aspects of the growth. Instead, it is
related to the complexity of the generated pattern.
\vspace{1cm} \\
PACs numbers: 05.40., 62.20M, 64.60., 81.40N
\end{abstract}
\newpage

A stimulating development in the analysis of complex systems has been
the hypothesis of {\it self organized criticality} \cite{btw,life}. The main
assumption is that such systems evolve towards a state where small
perturbations can give rise to changes (catastrophes) of all sizes.
This state describes the most unstable situation compatible with
some kind of equilibrium. The distribution of catastrophes, because of
its inherent scale invariance, is characterized by power laws.
In previous work, we have checked that this hypothesis is well satisfied
in systems which evolve into a steady state far from equilibrium
\cite{pra}.  The model we analyzed is Diffusion Limited Aggregation\cite{dla},
for which a comprehensive amount of work is already available.

In the present work we study the mechanisms through which scale invariant
noise is generated, as the complexity of the system is modified. The SOC
hypothesis is only applicable to situations which cannot be described by
simple, deterministic laws. When the behavior of the system is very
predictable, it will show only certain kinds of catastrophes, and not a power
law spectrum, as predicted by the SOC hypothesis.

To analyze in detail this question, we study a simple generalization of
the DLA model: the dielectric breakdown (DB) model with a scale invariant
growth law \cite{db}. As in
DLA, the growth is determined by a scalar field, $\phi$,
which obeys the Laplace equation outside the aggregate. The velocity of growth,
however, depends on the gradient of this field raised to some
power, $v \propto | \nabla \phi |^\eta$.
When $\eta = 1$, we recover DLA, and $\eta = 0$ describes the
Eden model. For $\eta \gg 1$, growth takes place
only in those points at the surface
of the aggregate where the field, $\nabla \phi$, has a maximum. Thus,
the system behaves in a simple and deterministic way. In this limit,
only narrow needles can grow,
with a fractal dimension close to one\cite{eta}.
It is clear the the change in the growth law does not introduce
any new scale in the system. Extensive calculations show that the
aggregates are, indeed, self similar. On the other hand, the noise
associated with the growth process can be reduced in a controlled
way, while keeping its intrinsic complexity\cite{noise}. Patterns
developed by this procedure are, at the same time, complex and
regular, reproducing the main features of dendritic growth.

We first investigate the changes in the noise spectrum as function
of $\eta$. The calculations reported below were obtained averaging results
for eight aggregates, grown in a circular cell of radius 108.
Preliminary results, as well as a detailed description of the
methods used are given in ref. \cite{granada}.

We define catastrophes following our previous work on DLA. We first
analyze aggregates of a given size. We calculate, once growth has
take place at certain site,
in how many lattice nodes outside the aggregate
the diffusion field changes above a given threshold, $\epsilon$. This
procedure allows us to obtain a definition of the spatial extent of
the rearrangement which follows a growth event. Alternatively, we
also study how many iterations are required to make the
diffussion field converge to its new equilibrium value, given an
overall error tolerancy in the calculations.
In this way, we can define the duration
of the catastrophe. We have made detailed calculations which show that
the SOC regime, as measured by its critical exponents, is
independent of the thresholds used. Also, modified definitions
of the size and duration of the catastrophes do not alter the
main results, that is, the existence of a power law distribution of
catastrophes, and only change slightly the exponents.
In our previous work \cite{pra,granada}, we have shown that there is a close
correspondence between the duration, and extension, of a
given catastrophe. Moreover, these quantities are uniquely
related to the strength of the field at the point where
the catastrophe (growth event) occured. This result can be intuitively
understood by noting that, whenever growth takes place at
a point where the field is large, a significant change in
the electrostatic potential around the aggregate follows,
because the new site in the aggregate has to be set at the
potential of the rest of the cluster. When the field,
that is, the potential drop between the aggregate and the
new site, was large, the rearrangement of the potential has also
to be large.

The study of the distribution of catastrophes in terms of the
electrical field at the surface of the aggregate allows us
to relate our work to the extensive study of the harmonic
measure of growing patterns\cite{har}. In addition, it is clear
that the noise associated with the growth of a given pattern
can be expressed in terms of the field distribution along
the perimeter of a static pattern. While the hypothesis of
SOC customarily describes the noise spectrum of a
dynamical process, better numerical accuracy, in our systems,
can be achieved be studying static distributions.

Our results for the Dielectric Breakdown Model are summarized
in figures 1 and 2. They show the distribution of the
values of the electric field as the aggregates grow (curves
labelled dynamical) and as function of location along the
perimeter of static aggregates (curves labelled statical).
Figure 1 shows the results obtained for $\eta = 0$ (Eden
model) and $\eta = 1$ (DLA). All curves show well defined
power laws over many decades. While in the Eden model the growth is
independent of the Laplacian field outside, we use
it as representative of low $\eta$ models.
In figure 2 we show results
for $\eta = 5$ and $\eta = 10$. There is no possibility of
defining a power law behavior in this regime. The values
of the electric field found in the dynamical simulations
tend to cluster at the upper range
of the curves. The fact that there is no longer dispersion
of the values of the fields (or catastrophe sizes) agrees
with the assumption that the aggregates have lost  their
intrinsic complexity. The growth sites, and their associated
fields are now highly predictable, despite the fact that
growth still takes place via an stochastic process\cite{needle}.
The observed values of the fields can be used to define
a scale in the system, which is not possible in the
regime where scale invariant noise appears.  The fields
tend to have their highest possible values, so that this scale
is the upper cutoff in the model, the
size of the aggregate. Thus, although we find a
qualitative change in the noise spectrum, it does
not happen in a similar way to phase transitions
in thermal equilibrium. If that were the case,
we expect a transition from a scale invariant
regime (characterized by one, or a line of
fixed points) to a situation where a new scale,
intermediate between the lattice size and the aggregate dimensions,
appears. On the other hand, we cannot rule out that the
system, in the high $\eta$ regime, evolves in bursts, growing first
at one tip, until that tip becomes blunt and growth there is arrested,
which leads to growth at another tip, and so forth. This possibility
resembles the first order phase transition model proposed as an
alternative to self organized criticality\cite{Nagel}.
\vskip 0.5cm
To complete or analysis of the role of complexity and external
noise in systems growing out of equilibrium,
we have analyzed the case
$\eta = 1$ with reduced noise. We use counters to achieve
the required degree of noise reduction\cite{noise}. Each possible
growth site has an attached variable. Each random choice of a
given site increases the value of that variable by one.
Real growth takes place when the variable exceeds a given threshold.
Once growth has occured, the variable at that point is reset
to zero. Noise reduction tends to increase the role of the
anisotropy of the underlying lattice, and the resulting patterns
resemble crystal growth.

In figure 3, we present the results obtained, using a threshold
of hundred for the noise reduction variable.
The field distributions are the average of three
aggregates. While the pattern is significantly different from
typical DLA aggregates, the noise spectrum, or the harmonic measure,
is not. The features in the pattern which are
due to the randomness in the growth process
dissapear, but the complexity of the system shows up
in the noise and related correlation functions.
The power law that we obtain is indistinguishable,
within our numerical accuracy, from the results for
standard DLA. The pattern shown in figure (3) is
highly symmetric and regular, but contains structure
on all scales. This fine structure is the cause of the
scale invariant noise spectrum shown in the
same figure.
Thus, even when the external noise is significantly reduced,
a broad band, scale invariant noise spectrum is generated.
Any small noise, or inhomogeneity in the initial
conditions, is amplified in the steady state.
This conclusion is rather general, and should be
applicable to a variety of experiments. We find
confirmation to it in the fact that, in dendritic growth,
the distribution of lengths of side branches seems
very irregular\cite{couder}.
That does not exclude that
the overall envelope coincides with simple, analytical, solutions.
\vskip 0.5cm
In conclusion, we have analyzed the circumstances under which
self organized critical behavior may appear. By studing models where
noise and intrinsic complexity can be modified, we show that SOC
is a feature directly related to the complexity of the evolving system.
Models which give rise to simple patterns do not exhibit scale
invariant noise, although their evolution includes noise terms.
On the other hand, growth laws which give rise to complex patterns,
show noise at all scales, irrespective of the stochasticity
included in the simulations.

We are thankful to V. Hakim and L. M. Sander for illuminating
conversations. Funding for this work has been provided
by the Comisi\'on Interministerial de Ciencia y Tecnolog{\'\i}a.

\section*{Figure Captions}
\ \\

{\bf Figure 1.} Histogram of electric fields at the perimeters of clusters
grown for $\eta = 0$ (upper graph) and $\eta = 1$ (lower graph) (thick curve).
Histogram of the values of fields at the sites where growth actually takes
place (thin curve). Inset shows the shape of one of the aggregates
for each value of $\eta$.

{\bf Figure 2.} As in figure 1, for $\eta = 5$ (upper graph) and $\eta =
10$ (lower graph).

{\bf Figure 3.}As in figures 1 and 2, but for clusters grown with
$\eta = 1$ and a noise reduction algorithm.

\begin{thebibliography}{99}
\bibitem{needle}A simple case which can be solved analytically is
when the aggregate can be modelled by a straight needle. Then, the
density of points with field $\epsilon$ is: $\Delta n / \Delta
\epsilon \propto \epsilon_{min}^2 / ( \epsilon^2 \sqrt{\epsilon^2
- \epsilon_{min}^2} )$, where $\epsilon_{min}$ is the value of the
field at the center of the needle.  This function has an
intrinsic scale set by $\epsilon_{min}$, which depends on the
length of the needle. Se also, B. Derrida and V. Hakim, preprint.
\end{thebibliography}
\end{document}